\documentclass[preprint]{aastex}

\usepackage{natbib}

\shorttitle{$\gamma$-ray Emission from  RGB\,J0710+591} 
\shortauthors{V. A. Acciari et al. (VERITAS collaboration)}

\begin{document}

\title{The Discovery of $\gamma$-ray Emission from the Blazar
  RGB\,J0710+591}
\author{
V.~A.~Acciari\altaffilmark{1},
E.~Aliu\altaffilmark{2},
T.~Arlen\altaffilmark{3},
T.~Aune\altaffilmark{4},
M.~Bautista\altaffilmark{5},
M.~Beilicke\altaffilmark{6},
W.~Benbow\altaffilmark{1},
M.~B{\"o}ttcher\altaffilmark{7},
D.~Boltuch\altaffilmark{8},
S.~M.~Bradbury\altaffilmark{9}
% ...
}

\altaffiltext{1}{Fred Lawrence Whipple Observatory, Harvard-Smithsonian Center for Astrophysics, Amado, AZ 85645, USA}
\altaffiltext{2}{Department of Physics and Astronomy, Barnard College, Columbia University, NY 10027, USA}
\altaffiltext{3}{Department of Physics and Astronomy, University of California, Los Angeles, CA 90095, USA}
\altaffiltext{4}{Santa Cruz Institute for Particle Physics and Department of Physics, University of California, Santa Cruz, CA 95064, USA}
\altaffiltext{5}{Physics Department, McGill University, Montreal, QC H3A 2T8, Canada}
\altaffiltext{6}{Department of Physics, Washington University, St. Louis, MO 63130, USA}
\altaffiltext{7}{Astrophysical Institute, Department of Physics and Astronomy, Ohio University, Athens, OH 45701}
\altaffiltext{8}{Department of Physics and Astronomy and the Bartol Research Institute, University of Delaware, Newark, DE 19716, USA}
\altaffiltext{9}{School of Physics and Astronomy, University of Leeds, Leeds, LS2 9JT, UK}
% ...

\begin{abstract}

  The high-frequency-peaked BL Lacertae object RGB\,J0710+591 was
  observed in the very-high-energy (VHE; E $>$ 100 GeV)
  wave band by the VERITAS array of
  atmospheric Cherenkov telescopes.  The observations, taken between
  December 2008 and March 2009 and totaling 22.1 hours, yield the
  discovery of VHE gamma rays from the source. RGB\,J0710+591 is
  detected at a statistical significance of 5.5 standard deviations
  ($5.5\sigma$) above the background, corresponding to an integral
  flux of $3.9 \pm 0.8 \times 10^{-12}~{\rm cm}^{-2}~{\rm s}^{-1}$
  (3\% of the Crab Nebula's flux) above 300 GeV.  The observed
  spectrum can be fit by a power law from 0.31 to 4.6 TeV with a
  photon spectral index of $2.69 \pm 0.26_{\rm stat} \pm
  0.20_{\rm sys}$.  These data are complemented by contemporaneous
  multiwavelength data from the {\it Fermi} Large Area Telescope, the
  {\it Swift} X-ray Telescope, the {\it Swift} Ultra-Violet and
  Optical Telescope and the Michigan-Dartmouth-MIT observatory.
  Modeling the broad-band spectral energy distribution with an
  equilibrium synchrotron self-Compton model yields a good statistical
  fit to the data. The addition of an external-Compton component to
  the model does not improve the fit nor bring the system closer to
  equipartition.  The combined {\it Fermi} and VERITAS data constrain
  the properties of the high-energy emission component of the source
  over four orders of magnitude and give 
  measurements of the rising and falling sections of the spectral
  energy distribution.
\end{abstract}

\keywords{gamma rays: galaxies --- BL Lacertae objects: individual
  (RGB\,J0710+591, EXO\,0706.1+5913, VER\,J0710+591)}

\section{Introduction}

RGB\,J0710+591 was originally discovered by HEAO A-1
\citep{Wood:1984yq}; it is a BL Lac object located at a redshift of $z
= 0.125$ \citep{Giommi:1991eu}.  Hubble Space Telescope observations
of this blazar show a fully resolved diskless galaxy that can be
modeled using an elliptical profile with a nuclear point source
\citep{Scarpa:2000gd}.  In addition, RGB\,J0710+591 is found to have
an optical `compact companion' located at a distance of $1.4''$ (3.5
kpc) from the core of the active galactic nucleus (AGN). The nature of
this companion is unknown, but it could be a foreground star or an
unresolved faint galaxy.  It is, however, too luminous for a globular
cluster \citep{Falomo:2000lr} and is not seen in radio
\citep{Giroletti:2004lr}.

Blazars are characterized by spectral energy distributions (SEDs)
consisting of two peaks, the low energy peak arising from synchrotron
emission and the high energy peak from either leptonic or hadronic
interactions \citep[for a review of the emission mechanisms in
blazars, see][]{Bottcher:2007eu}.  Depending on the location of the
low energy peak, BL Lacs are classified as being high-frequency-peaked
(HBL, $\nu_{peak} \sim 10^{16-18}$ Hz), intermediate-frequency-peaked
(IBL, $\nu_{peak} \sim 10^{15-16}$ Hz), or low-frequency-peaked (LBL,
$\nu_{peak} \sim 10^{13-15}$ Hz).  The location of the low energy SED
peak of RGB\,J0710+591 (identified as EXO\,0706.1+5913 in the
following reference) is higher than $\sim\!10^{18}$ Hz
\citep{Nieppola:2006pb}, clearly identifying it as an HBL.

RGB\,J0710+591 is a well known HBL featured in many catalogs, but it
was not detected in the high-energy (HE; E $>$ 100 MeV) band by the
EGRET instrument on the Compton Gamma-Ray Observatory.  In the
very-high-energy (VHE; E $>$ 100 GeV) band, the Whipple 10-m imaging
atmospheric Cherenkov telescope (IACT) reported several flux upper
limits for observations taken in different epochs: (0.91, 1.69, 3.58,
and 4.29) $\times10^{-12}~{\rm cm}^{-2}~{\rm s}^{-1}$ (99.9\%
confidence level) above (350, 350, 500 and 400 GeV) respectively,
corresponding to (8.7\%, 16.1\%, 52.4\% and 45.9\% ) of the Crab
Nebula's flux \citep{Horan:2004yo}.  The HEGRA IACT system reported a
flux upper limit above 1.08 TeV of $0.91 \times 10^{-12}~{\rm
  cm}^{-2}~{\rm s}^{-1}$ (99\% confidence level) corresponding to 6\%
of the Crab Nebula's flux \citep{Aharonian:2004lq}.

There are several compelling reasons to study RGB\,J0710+591 in the
MeV to TeV energy range.  Due to the hardness of its X-ray spectrum,
the high energy peak is expected to be located in an ideal location
for HE and VHE observations.  There are both leptonic
\citep{Ghisellini:1996oq,Bottcher:2002mb} and hadronic models
\citep{Aharonian:2000vh,Mucke:2003hb} that explain the
photon-production mechanism of the high energy peak.  The continuous
measurement of the SED from MeV to TeV energies is vital to constrain
certain aspects of the emission models and to perhaps exclude some of
them entirely.  For example, some leptonic models appear harder at HEs
than hadronic models. At VHE energies, photons interact via
pair-production with the infrared component of the extragalactic
background light (EBL) \citep{Gould:1967nx}.  This causes a decrease
in the observed flux and a softening of the observed spectrum.  Thus,
extragalactic objects at moderate redshifts with reasonably hard VHE
spectra are ideal candidates to indirectly study the history of the
EBL and its SED, particularly in the optical to far infrared
\citep[see, e.g.,][]{Aharonian:2007nx}.

VHE $\gamma$-ray emission from RGB\,J0710+591 was discovered by
VERITAS during the 2008-2009 observing season \citep{ong:2009rgb}.
Following this, the VERITAS collaboration initiated a multiwavelength
(MWL) observation campaign that included the participation of the {\it
  Fermi} Large Area Telescope (LAT) in HE $\gamma$-rays, the {\it
  Swift} X-ray Telescope (XRT) in X-rays and the {\it Swift}
Ultra-Violet and Optical Telescope (UVOT) and the
Michigan-Dartmouth-MIT (MDM) observatory in optical.  The resulting
MWL SED spans thirteen decades in energy.

\section{VERITAS Observations}

VERITAS is an array of four 12-m IACTs located in southern Arizona at
the F.~L.~Whipple Observatory \citep{Weekes:2002pi,Holder:2008rm}.
VERITAS has an energy resolution of $\sim\!15\%$ between 100 GeV and
30 TeV, an angular resolution of $\sim\!0.1^\circ$ (68\% containment)
per event, and a field of view (FoV) of $3.5^\circ$.  VERITAS can
detect a source with a flux of 1\% of the Crab in $< 50$ hours, while
a $5\%$ Crab Nebula flux source is detected in $\sim\!2.5$ hours.
These specifications are valid for the data taken during this
campaign; the VERITAS array was subsequently upgraded, resulting in a
significant improvement in sensitivity \citep{Perkins:2009}.

\begin{figure}
\plotone{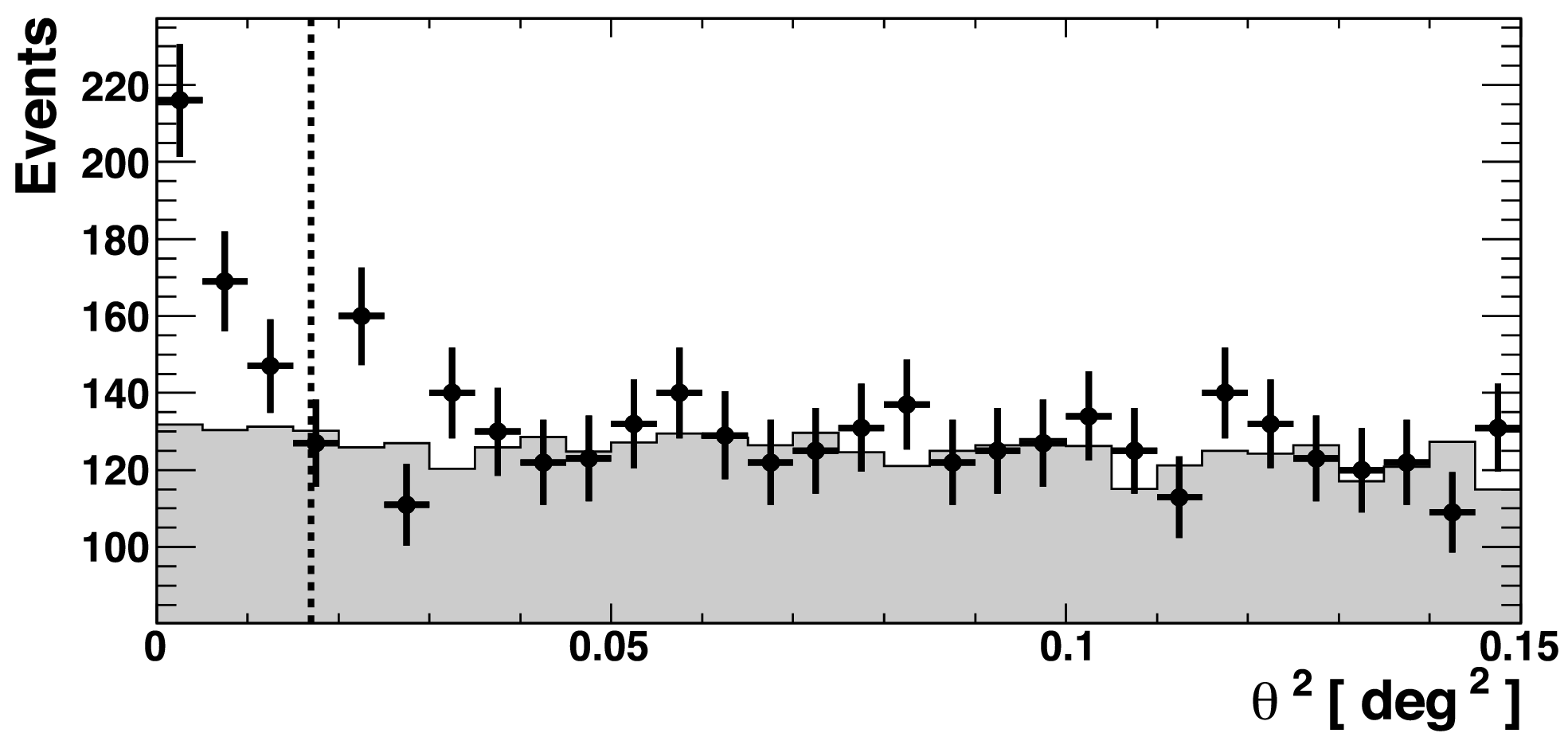}
\caption{\label{fig:theta2}The distribution of $\theta^2$ for
  on-source VHE events (point
  (shaded region) from observations of RGB\,J0710+591. $\theta$ is the
  distance from the source location to the event location.  The dashed
  line represents the location of the cut on $\theta^2$ applied to the
  data.  The $\theta^2$ cut retains events which lie to the
    left of the line.}
\end{figure}

\begin{figure}
\plotone{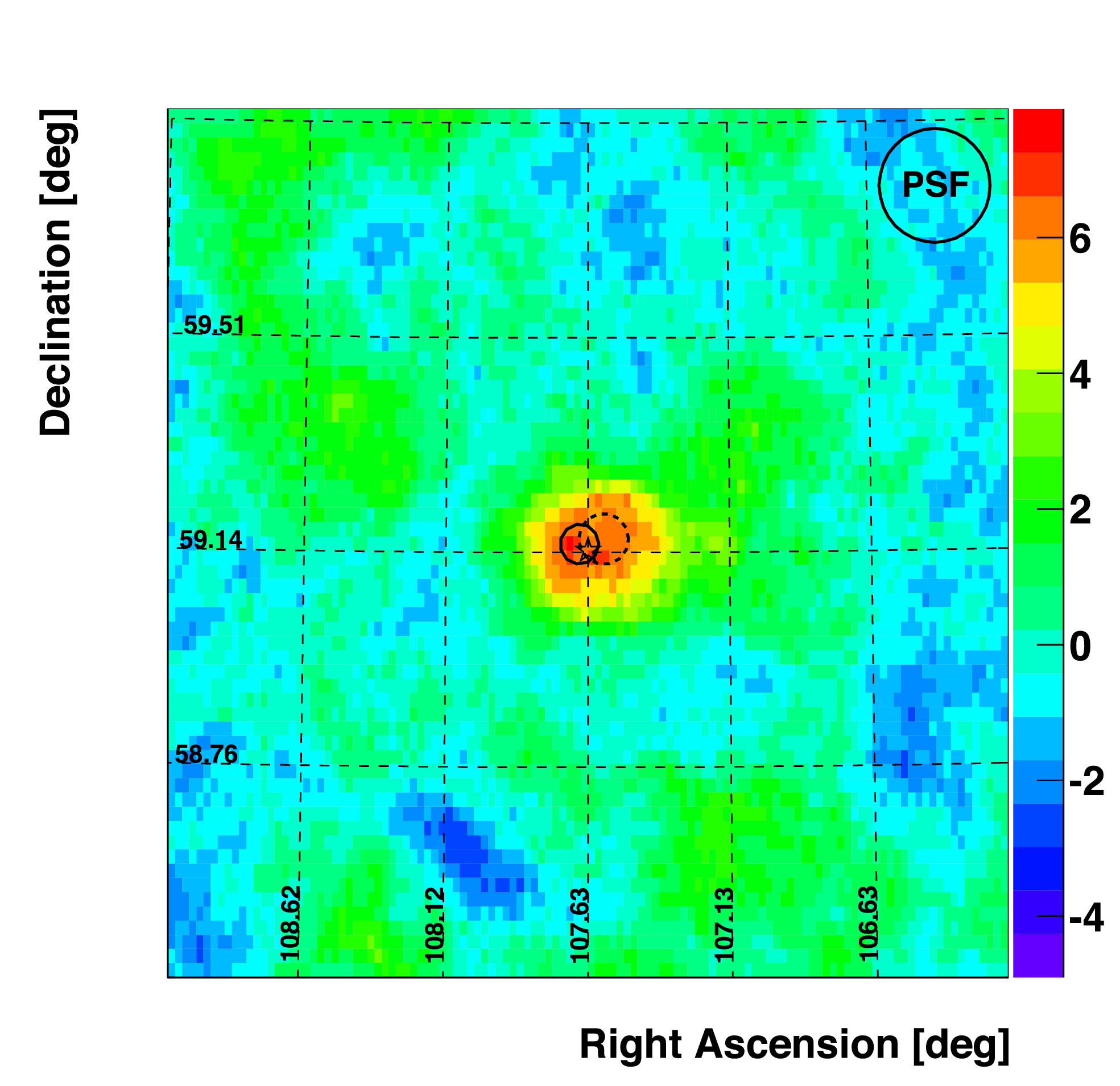}
\caption{\label{fig:skymap}Smoothed VHE significance map of
  RGB\,J0710+591 from the VERITAS observations.  The VLA
  location of the blazar is shown as a star.  The solid circle
  is centered at
    the best fit location of the excess emission in the VERITAS data;
    the radius of the circle is the sum of the statistical and
    systematic uncertainties. The dashed circle is the best fit plus a
  95\% containment radius of $2.7'$ to the {\it Fermi} LAT data.  Both
  locations are consistent with the VLA position of RGB\,J0710+591.
  The VERITAS point spread function is shown in the top
    right corner.}
\end{figure}

RGB\,J0710+591 was observed with VERITAS from December 2008 to March
2009 during astronomical darkness and partial moonlight for a total
time of 34.9 hours.  Approximately 30\% of these observations are
rejected due to poor weather or hardware-related problems with the
majority rejected due to weather, resulting in 22.1 hours of exposure
(livetime corrected).  The average zenith angle of the observations is
$30^\circ$.  The VERITAS data are calibrated and cleaned as described
in \cite{Daniel:2008lr}.  After calibration, several noise-reducing
selection criteria (cuts) are made, including rejecting those events
where only the two closest-spaced telescopes participated in the
trigger.  The VERITAS standard analysis consists of stereo
parametrization using a moment analysis \citep{Hillas:1985ta} and
following this, scaled parameters are used for event selection
\citep{Aharonian:1997rm,Krawczynski:2006ts}.  This selection is based
on a set of cuts that are {\it a priori} optimized using data taken on
the Crab Nebula where the $\gamma$-ray excess has been scaled by 0.01
to account for weaker sources like RGB\,J0710+591. VERITAS observed
RGB\,J0710+591 in ``wobble'' mode, where the center of the FoV is
offset by $0.5^\circ$ from the blazar, to allow for simultaneous
source and background measurements \citep{Berge:2007ud} The reflected
region model \citep{Aharonian:2001fk} is used to calculate the
background events from several off-source regions.  A total of 576
on-source events are observed from the direction of RGB\,J0710+591 and
3890 are seen in the background regions. Depending on the pointing
direction either eight or nine background regions are used resulting
in a final $\alpha$ of 0.115 ($\alpha$ is the ratio of the areas of
the on and off-source regions).  This results in an excess of 129
events. This excess has a statistical significance of 5.5 standard
deviations ($\sigma$) and corresponds to an integral flux of $3.9 \pm
0.8 \times 10^{-12}~cm^{-2}~s^{-1}$ (3\% of the Crab Nebula's flux)
above 300 GeV.  There is no evidence for variability during these
observations (a fit of the lightcurve to a constant flux yields a
reduced chi-squared value of 0.21). The excess is point-like and is
located at $07^{\rm h}~10^{\rm m}~26.4^{\rm s}\pm2.4^{\rm s}_{\rm
  stat}$, $59^\circ~09'~00''\pm36''_{\rm stat}$ (J2000, see Figure
\ref{fig:skymap}; the systematic uncertainty in the pointing is
$90''$) which is $49''$ away from the location of RGB\,J0710+591
($07^{\rm h}~10^{\rm m}~30.07^{\rm s},59^\circ~08'~20.5''$) measured
by the Very Large Array at 5 GHz \citep{Laurent-Muehleisen:1999kx}.
The VERITAS source name is VER\,J0710+591.  The $\theta^2$
distributions for the on-source and background regions are shown in
Figure \ref{fig:theta2} ($\theta$ is the distance between the source
location and the location of an event) . The photon energy spectrum
can be fit by a simple power-law ($dN/dE = I (E/E_0)^{-\Gamma}$, $E_0$
= 1 TeV) with photon index $\Gamma_{\rm VHE} = 2.69 \pm 0.26_{\rm
  stat} \pm 0.20_{\rm sys}$ (see Table \ref{tab:spectrum} for the
spectral points).  The fit yields a $\chi^2$ per degree of freedom of
$\chi^2/dof = 1.3/3$.
 
\begin{table}
\begin{center}
  \caption{The VERITAS spectral bins. `E$_{low}$' and
      `E$_{high}$' are the low and high energy bounds, `Flux' is the
      differential flux, `Excess' is the number of excess events and
      `Sig.' is the significance of the bin.\label{tab:spectrum}}
\begin{tabular}{cccccc}
  \tableline\tableline
  E$_{low}$ & E$_{high}$ & Flux & Flux Error & Excess & Sig. \\
  $[{\rm TeV}]$    & [TeV]  & [cm$^{-2}$~TeV$^{-1}$~s$^{-1}$]& [cm$^{-2}$~TeV$^{-1}$~s$^{-1}$] & & [$\sigma$] \\
  \tableline
  0.31 & 0.53 & $8.9\times10^{-12}$  & $3.6\times10^{-12}$ & 29.9 & 2.7 \\
  0.53 & 0.92 & $2.4\times10^{-12}$  & $7.7\times10^{-13}$ & 32.9 &3.5 \\
  0.92 & 1.6 & $7.3\times10^{-13}$  & $2.2\times10^{-13}$ & 24.3 & 3.9 \\
  1.6 & 2.7 & $8.6\times10^{-14}$  & $5.6\times10^{-14}$ & 6.07 & 1.7 \\
  2.7 & 4.6 & $3.1\times10^{-14}$  & $2.0\times10^{-14}$ & 4.25 & 2.0 \\
  4.6 & 7.9 & $7.4\times10^{-15}$  & {\it 99\% upper limit} & -0.26 & -0.23 \\ 
  \tableline
\end{tabular}
\end{center}
\end{table}

\section{{\it Fermi} Observations}

The {\it Fermi} LAT is a pair-conversion gamma-ray detector sensitive
to photons in the energy range from below 20 MeV to more than 300 GeV
\citep{Atwood:2009rm}. The LAT data for this analysis were taken
between 4 August 2008 and 4 August 2009. The data are analyzed using
the standard likelihood tools distributed with the Science Tools
v9r15p3 package available from the {\it Fermi} Science Support Center
\footnote{http://fermi.gsfc.nasa.gov/ssc/}. Only events having the
highest probability of being photons, those in the ``diffuse'' class,
are used. To limit contamination by Earth albedo gamma rays which are
produced by cosmic rays interacting with the upper atmosphere, only
events with zenith angles $< 105^\circ$ are selected. Events with
energy between 100 MeV and 300 GeV and within a circular region of
$10^\circ$ radius centered on the VLA coordinates of RGB\,J0710+591
are analyzed with an unbinned maximum likelihood method
\citep{Cash:1979yq,Mattox:1996vn}. The background emission is modeled
using a Galactic diffuse emission model and an isotropic
component.\footnote{http://fermi.gsfc.nasa.gov/ssc/data/access/lat/BackgroundModels.html}
The fluxes are determined using the post-launch instrument response
functions P6\_V3\_DIFFUSE.

A point source is detected with a significance of more than 8 standard
deviations. The best-fit position ($07^{\rm h}~10^{\rm m}~37^{\rm s}$,
$59^\circ~09'~32''$) has a 95\% containment radius of $2.7'$ and is
statistically consistent with the VLA coordinates of RGB\,J0710+591
(see Figure \ref{fig:skymap}). The highest energy photon associated
with the source has an energy of 75 GeV. The time-averaged {\it Fermi}
spectrum is fit by a power-law for which $I = 1.43 \pm 0.35_{\rm stat}
\pm 0.10_{\rm sys} \times 10^{-14} ~ {\rm cm}^{-2} s^{-1}~{\rm
  MeV}^{-1}, \Gamma_{\rm HE} = 1.46 \pm 0.17_{\rm stat} \pm 0.05_{\rm
  sys}$, and $E_0 = 8.8$ GeV is the energy at which the correlation
between the fit values of I and $\Gamma_{\rm HE}$ is minimized. The
spectral index reported in the 1FGL catalog \citep{Abdo:2010aa} for
this source is $1.28 \pm 0.21_{\rm stat}$. The discrepancy between
that result and the result reported here can be explained by the
different energy range considered in the likelihood model of the
source: 100 MeV - 100 GeV for the catalog analysis, 100 MeV - 300 GeV
for the present analysis.  The lack of photons above 75 GeV carries
information that is important for the fit results.  No
evidence for variability is detected.

\section{{\it Swift} Observations}

A total of seven {\it Swift} \citep{Gehrels:2004oq} XRT and UVOT
observations of RGB\,J0710+591 were performed between 20 February 2009
and 2 March 2009 (observation identifications 00031356001 through
00031356007). The total XRT observation time is 15.2 ks while the
total UVOT observation time is 14.8 ks. The {\it Swift} XRT data are
analyzed with HEAsoft 6.5 using the most recent calibration files. The
first three XRT observations in photon-counting mode have source count
rates of $\sim$0.7--1.5 counts s$^{-1}$, resulting in significant
photon pile-up in the core of the point-spread function, which occurs
for rates $>$0.5 counts s$^{-1}$. The photon pile-up is avoided by
using an annular region with inner radius ranging from 2 to 7 pixels
and an outer radius of 30 pixels ($47.2''$) to extract the source
counts. Background counts are extracted from a 40 pixel radius circle
in a source-free region. For the last four XRT observations in
windowed-timing mode, source counts are extracted from a rectangular
region of 40 by 20 pixels high, with background counts extracted from
a nearby source-free rectangular region of equivalent size. Evidence
for marginal flux variability between the {\it Swift} XRT observations
is indicated by a probability of 0.004 for a fit of a constant
flux. The time-averaged XRT data can be fit with an absorbed power-law
model where the hydrogen column density N$_{\rm{H}}$ is fixed at $4.4
\times 10^{20}$ cm$^{-2}$ \citep{Kalberla:2005zl}, resulting in a
combined spectrum with photon index of $1.86 \pm 0.01$, with a
$\chi^{2}$/dof $=$ 371 / 303. This hard spectrum is consistent with a
synchrotron peak above 10 keV, as can be inferred from the radio to
X-ray spectral energy distribution (SED) generated from archival data,
including an EXOSAT detection \citep{Giommi:1995dq}.

\begin{figure}
\plotone{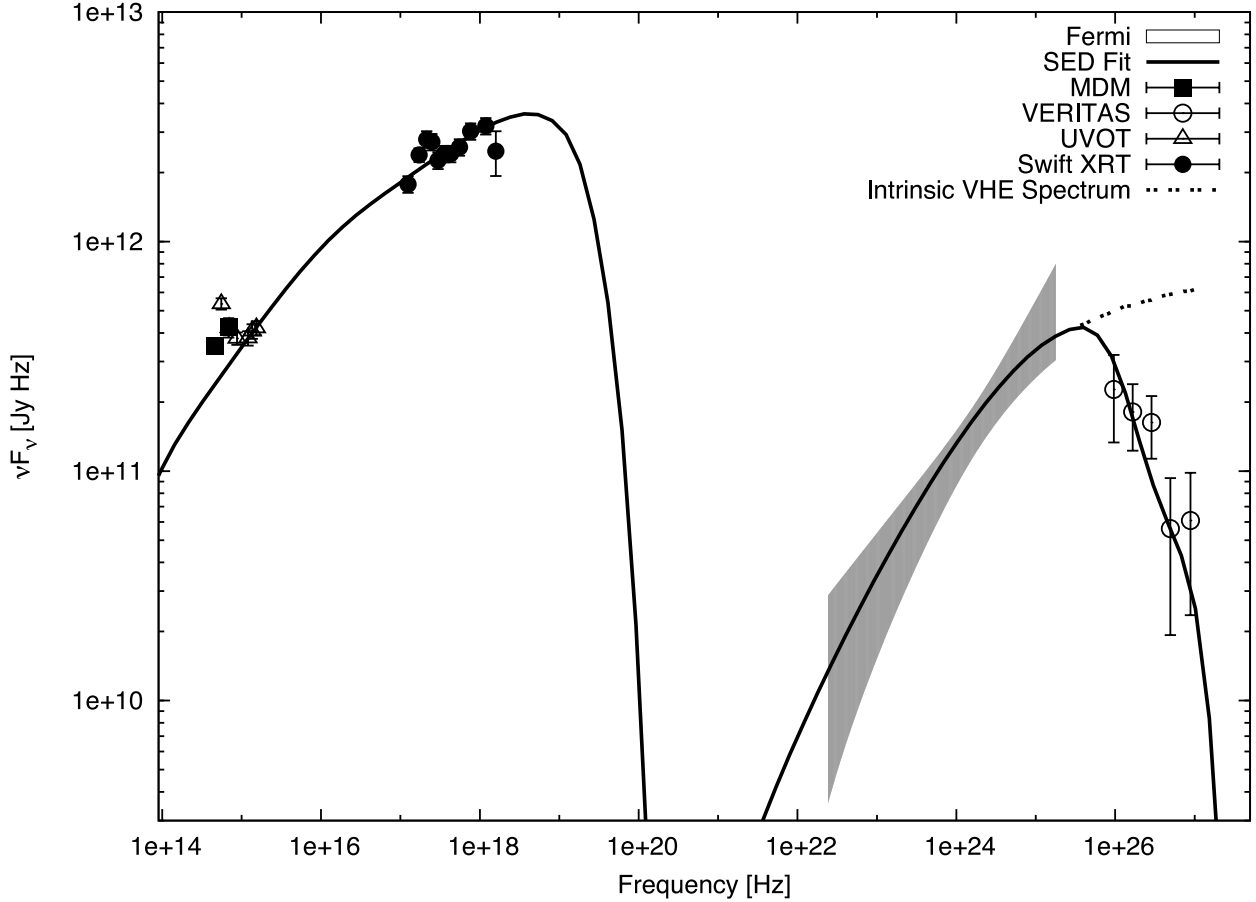}
\caption{\label{fig:sed}Broad-band SED of RGB\,J0710+591.  The closed
  squares are R- and B-band data from the MDM 1.3m telescope, the open
  triangles are from the {\it Swift} UVOT, closed circles are from the
  {\it Swift} XRT, the shaded area is the {\it Fermi} LAT measurement
  and the open circles are the measured VERITAS data.  The broad-band
  spectral model is the equilibrium SSC version from
  \citet{Bottcher:2002mb} with a cooling break energy occurring around
  $1.3\times10^{16}$ Hz.  Note that the optical measurements (MDM and
  {\it Swift} {\it u}, {\it b} and {\it v} bands) are affected by
  emission from the host galaxy and are thus not used in the SED
  fit. The SED model curve includes EBL absorption using the model of
  \citet{Franceschini:2008il} and the dashed line is the
    de-absorbed VHE spectrum.}
\end{figure}

{\it Swift} UVOT \citep{Roming:2005ud} data are presented in the six
bands of {\it v}, {\it b}, {\it u}, {\it uvm}1, {\it uvm}2 and {\it
  uvw}2. The UVOTSOURCE tool is used to extract counts, correct for
coincidence losses, apply background subtraction, and calculate the
source flux \citep{Poole:2008kl}. The standard $5''$ radius source
aperture is used, with a $20''$ background region. The source fluxes
are extinction corrected using the dust maps of
\citet{Schlegel:1998kx}.  Since the stacked UVOT images are not deep
enough to sample the host galaxy profile, no attempt is made to
correct for the contribution of the host galaxy and this can be seen
as an increase in flux in the optical portion of the SED (Figure
\ref{fig:sed}) starting with the {\it u}-band.

\section{MDM Observations}

The observations of RGB\,J0710+591 taken from February 19 through 24,
2009 with the MDM Observatory 1.3m
telescope\footnote{http://www.astro.lsa.umich.edu/obs/mdm/} in R and B
band are bias-subtracted and flat-fielded using the Image Reduction
and Analysis Facility (IRAF) \citep{Tody:1993vf,Tody:1986uk}.
Instrumental magnitudes are extracted using the IRAF DAOPHOT package
and calibrated using four comparison stars in the field, with
comparison star magnitudes from the Naval Observatory Merged
Astrometric Dataset (NOMAD) catalog \citep{Zacharias:2004kx}, and
extinction coefficients $A_B = 0.166, A_R = 0.106$ from the NASA/IPAC
Extragalactic Database
(NED)\footnote{http://nedwww.ipac.caltech.edu/}.

The host galaxy of RGB\,J0710+591 makes a significant contribution to
the optical fluxes measured by MDM.  The host galaxy subtractions for
this BL Lac were calculated previously in \citet{Hyvonen:2007nx} and
these subtractions are used to correct for the host galaxy
contribution to the flux measurements.  The resulting extinction- and
host-galaxy-corrected optical flux in the R-band ($4.68\times10^{14}$
Hz) is $3.51 \pm 0.07 \times10^{11}$ Jy Hz and in the B-band
($6.85\times10^{14}$ Hz) is $4.23 \pm 0.41 \times 10^{11}$ JyHz.  Even
with the host galaxy corrections, the MDM data are still higher than
the UV measurements made by the {\it Swift} UVOT which are dominated
by the nucleus.  Thus, the MDM and {\it Swift} optical data in the
{\it u}, {\it b} and {\it v} bands still have contributions due to the
host galaxy and should be considered upper limits. No substantial
optical variability is seen over the course of the five day
observation period.

\section{Modeling and Discussion}

The observed broad-band SED of RGB\,J0710+591 is shown in Figure
\ref{fig:sed}. The data, excluding the MDM and Swift {\it u}, {\it b}
and {\it v} bands which are affected by contributions from the host,
are fit with a purely leptonic synchrotron self-Compton (SSC) model.
Specifically, an equilibrium version of the model of
\cite{Bottcher:2002mb}, as described in \cite{Acciari:2009wc} is
used. The best-fit parameters are listed in Table \ref{tab:model}.
The VHE data in the plot are the measured values and the model
includes the effects of intergalactic $\gamma-\gamma$ absorption using
the EBL models of \citet{Franceschini:2008il} and the reported
redshift of $z = 0.125$.  De-absorbing the VHE data using the same
models results in a de-absorbed spectral index of $\Gamma_{\rm VHE}
\sim 1.85$.

Many of the model parameters are not constrained by the
observations. In particular, the choice of the bulk Lorentz factor
($\Gamma_{\rm bulk}$) and observing angle ($\theta_{obs}$) are
somewhat arbitrary and similarly good fits can be obtained using other
values. Evidence from portions of the radio structure is supportive of
a small observing angle which corroborates the small value obtained
here.  However, the bulk Lorentz factor derived from these same data
is lower \citep{Giroletti:2006lr}. Regardless, the synchrotron portion
of the SED always turns over far below the X-ray regime. Therefore, in
order to obtain the observed hard X-ray spectrum, a very hard
injection index (q = 1.5) has to be assumed. This poses challenges to
standard (1st-order) Fermi acceleration models \citep{Stecker:2007kx},
and it might indicate second-order Fermi acceleration or other modes
of acceleration (e.g., shear at boundary layers). The fit requires a
rather low magnetic field of $B = 0.036 {\rm G}$ , far below
equipartition. This cannot be remedied even with the choice of a
higher Doppler factor.

Additionally, an external-Compton (EC) component was added to the
model to see if the resulting fit would improve, or if the system
would be closer to equipartition.  However, in order to produce an EC
peak at sufficiently high frequencies, a very low magnetic field is
needed.  Thus, the inclusion of an EC component does not lead to an
overall improvement of the fit statistically, or with respect to
equipartition.

\begin{table}
\begin{center}
  \caption{The equilibrium
    model parameters for the purely leptonic SSC model that was used
    to fit the SED of RGB\,J0170+591.  Details of the model can be
    found in the text.\label{tab:model}}
\begin{tabular}{cc}
\tableline\tableline
Parameter & Value \\
\tableline
$\gamma_{min}$ & $6\times10^4$ \\
$\gamma_{max}$ & $2\times10^6$ \\
$e^-$ Injection spectral index &1.5 \\
Escape time parameter & $\eta_{esc}$ = 100\tablenotemark{a} \\
Magnetic field at $z_0$ & 0.036 G\\
Bulk Lorentz factor & $\Gamma_{\rm bulk}$ = 30\\
Blob radius & $2\times10^{16}$ cm\\
$\theta_{obs}$ & 1.91\tablenotemark{b} degrees\\
Redshift & z = 0.125 \\
$L_e$ (jet) & $4.49\times10^{43}$ erg/s\\
$L_B$ (jet) & $1.75\times10^{42}$ erg/s\\
$L_B$/$L_e$ & $3.90\times10^{-2}$\\
\tableline
\end{tabular}

\tablenotetext{a}{$t_{esc} = \eta_{esc}*R/c$}
\tablenotetext{b}{equivalent to the superluminal angle}

\end{center}
\end{table}

With a photon index $\Gamma_{\rm VHE} = 2.69$, RGB\,J0710+591 is one
of the hardest VHE blazars detected to date.  Furthermore, out of the
19 blazars detected by both VHE instruments and {\it Fermi},
RGB\,J0710+591 has the hardest HE photon index \citep{Fegan:2009qq}
and \citet{Nieppola:2006pb} found that RGB\,J0710+591 has one of the
highest low energy SED peaks out of a sample of over 300 BL Lacs and
place it in a list of 22 ultra-high-frequency-peaked
candidates. Nevertheless, the observed SED is well fit by a purely
leptonic SSC model without the need to include an external component.

Observations of the similarly hard ($\Gamma_{\rm VHE} = 2.50$) and
distant ({\it z} = 0.1396) blazar 1ES\,0229+200
\citep{Aharonian:2007nx} indicated that the EBL density in the mid-IR
band was close to the lower limits measured by Spitzer.  These
constraints on the EBL are based on the assumption that the intrinsic
spectrum of blazars must have a spectral index of 1.5 or larger.
Thus, if blazars like 1ES\,0229+200 and RGB\,J0710+591 are unusual
objects this assumption might not be valid because their intrinsic
spectral indices could be smaller than 1.5. However, the modeling of
RGB\,J0710+591 and other similar blazars indicates that hard-spectrum
VHE blazars might not be exotic (i.e. having an intrinsic spectral
index less than 1.5) and are part of the general blazar
population. Therefore, using VHE observations of hard spectrum blazars
to constrain EBL models as done in \cite{Aharonian:2007nx} continues
to be a valid undertaking.

\acknowledgements

The VERITAS collaboration acknowledges support from the
U.S. Department of Energy, the U.S. National Science Foundation and
the Smithsonian Institution, by NSERC in Canada, by Science Foundation
Ireland and by STFC in the UK.  This research has made use of the
NASA/IPAC Extragalactic Database (NED) which is operated by the Jet
Propulsion Laboratory, California Institute of Technology, under
contract with the National Aeronautics and Space Administration.

The \textit{Fermi} LAT Collaboration acknowledges generous ongoing
support from a number of agencies and institutes that have supported
both the development and the operation of the LAT as well as
scientific data analysis.  These include the National Aeronautics and
Space Administration and the Department of Energy in the United
States, the Commissariat \`a l'Energie Atomique and the Centre
National de la Recherche Scientifique / Institut National de Physique
Nucl\'eaire et de Physique des Particules in France, the Agenzia
Spaziale Italiana and the Istituto Nazionale di Fisica Nucleare in
Italy, the Ministry of Education, Culture, Sports, Science and
Technology (MEXT), High Energy Accelerator Research Organization (KEK)
and Japan Aerospace Exploration Agency (JAXA) in Japan, and the
K.~A.~Wallenberg Foundation, the Swedish Research Council and the
Swedish National Space Board in Sweden.

Additional support for science analysis during the operations phase is
gratefully acknowledged from the Istituto Nazionale di Astrofisica in
Italy and the Centre National d'\'Etudes Spatiales in France.

\end{document}